\documentclass[%
 aip,
 pof,
 amsmath,amssymb,
 preprint,%
]{revtex4-1}

\usepackage{graphicx}
\usepackage{dcolumn}
\usepackage{bm}
\usepackage{url}
\usepackage[T1]{fontenc}
\usepackage[utf8]{inputenc}
\usepackage{mathptmx}
\usepackage{etoolbox}

\makeatletter
\def\@email#1#2{%
 \endgroup
 \patchcmd{\titleblock@produce}
  {\frontmatter@RRAPformat}
  {\frontmatter@RRAPformat{\produce@RRAP{*#1\href{mailto:#2}{#2}}}\frontmatter@RRAPformat}
  {}{}
}%
\makeatother

\newcommand{\eps}{\epsilon}
\newcommand{\dd}{\mathrm{d}}
\newcommand{\vect}[1]{\bm{#1}}

\newcommand{\maybeincludegraphics}[2][]{%
  \IfFileExists{#2.pdf}{\includegraphics[#1]{#2.pdf}}{%
  \IfFileExists{#2.png}{\includegraphics[#1]{#2.png}}{%
  \IfFileExists{#2.jpg}{\includegraphics[#1]{#2.jpg}}{%
  \fbox{\parbox{0.8\textwidth}{\centering Figure file not found: \detokenize{#2}}}}}}%
}
\newcommand{\Rey}{\mathrm{Re}}
\newcommand{\St}{\mathrm{St}}

\begin{document}

\title{Finite-Time Relaxation of Inertial Particle Clustering in Non-Equilibrium Turbulence}

\author{Taketo TOMINAGA}
\email{tominaga.t.2d38@m.isct.ac.jp}
\affiliation{School of Engineering, Department of Mechanical Engineering, Institute of Science Tokyo}

\author{Ryo ONISHI}
\affiliation{Supercomputing Research Center, Institute of Integrated Research, Institute of Science Tokyo}

\date{\today}

\begin{abstract}
Inertial particles in turbulence form clusters, which strongly affect particle collisions and transport properties.
Clustering models based on statistically stationary turbulence implicitly assume the instantaneous-equilibrium approximation when applied to time-varying non-equilibrium turbulence.
However, the validity of this approximation remains unclear.
In this study, the temporal response of inertial particle clustering in non-equilibrium turbulence was investigated using direct numerical simulation of homogeneous isotropic turbulence with unsteady forcing.
Periodic responses of the flow and clustering intensity were evaluated by varying the forcing period.
The flow showed non-equilibrium scaling for all forcing periods.
The relationship between instantaneous energy dissipation rate and clustering intensity showed hysteresis exceeding statistically stationary fluctuations when the forcing period exceeded several large-eddy turnover times.
For the particles with the largest inertia, clustering intensity took values of 0.80 and 1.56 times the reference value at the same instantaneous energy dissipation rate.
This demonstrates that the instantaneous-equilibrium approximation is not appropriate under such conditions.
A linear relaxation model was constructed from transient responses, in which clustering intensity approaches the instantaneous-equilibrium value with a finite relaxation time.
The relaxation time scaling was identified as $\tau_g = 1.0 T_\mathrm{e}(t)\,\mathrm{St}(t)^{0.40}$, where $T_\mathrm{e}(t)$ and $\mathrm{St}(t)$ are the instantaneous large-eddy turnover time and Stokes number.
The model reduced the maximum relative error from 49\% to 10\% for the particles with the largest inertia and from 76\% to 22\% in an independent validation case.
These results demonstrate that finite-time relaxation improves prediction accuracy for clustering intensity in non-equilibrium turbulence.
\end{abstract}

\maketitle

\section{Introduction}

Inertial particles in turbulence appear in a wide range of geophysical and engineering applications, such as cloud turbulence, aerosol transport, spray combustion, and solid-particle transport.
Inertial particles are not uniformly distributed in turbulence and can form highly non-uniform spatial distributions, known as preferential concentration \cite{maxey1987the,squires1991measurements}.
Such non-uniform particle distributions are referred to as clustering and strongly affect particle collisions and transport properties \cite{sundaram1997collision,wang2000statistical}.

In cloud microphysics, the enhancement of cloud droplet collisions by turbulence has been widely studied as one mechanism for the rapid formation of warm rain \cite{falkovich2007sling,grabowski2009diffusional,onishi2015lagrangian}.
In particular, inertial particle clustering increases the collision kernel by increasing the radial distribution function (RDF) $g(r = R)$ at the contact distance $R = 2r_{\mathrm{p}}$, where $r_{\mathrm{p}}$ is the particle radius \cite{wang2000statistical,onishi2015lagrangian}.
To quantify this effect, $g(r = R)$ and the collision kernel have been modeled based on direct numerical simulation (DNS) and theoretical analyses \cite{saffman1956on,wang2000statistical,ayala2008effects,wang2008turbulent,onishi2016reynolds}.
In addition, the effect of turbulence on cloud droplet growth has been investigated by incorporating these collision kernels into the stochastic collision--coalescence equation \cite{xue2008growth,onishi2015lagrangian,onishi2016reynolds}.

These models are based on clustering statistics in statistically stationary turbulent flows.
Therefore, when they are applied to non-equilibrium turbulent flows in which the turbulent state varies in time, it is implicitly assumed that $g(r = R)$ instantaneously changes to the equilibrium value corresponding to the turbulent state at that time.
However, the validity of this assumption remains unclear.
Indeed, Zapata et al. \cite{zapata2024turbulence} studied particle clustering in non-equilibrium turbulence that spontaneously emerges under statistically stationary counter-rotating-column forcing.
They reported that particle clustering responds to the turbulent state with a time lag and exhibits hysteresis.
Nevertheless, the hysteresis of clustering intensity and its relaxation time have not been quantitatively characterized.

In this study, the temporal response of inertial particle clustering in non-equilibrium turbulence is investigated using DNS of homogeneous isotropic turbulence with unsteady forcing, in which the energy injection rate varies in time.
First, the response to periodic unsteady forcing is analyzed, and hysteresis in the relationship between the energy dissipation rate and clustering intensity is evaluated.
Furthermore, a linear relaxation model is constructed, in which clustering intensity approaches the instantaneous-equilibrium value with finite-time relaxation.
The relaxation time scaling included in the model is identified from transient-response DNS data obtained by switching the energy injection rate only once.
The predictive performance of the obtained model is validated using independent DNS data.
This study provides a basis for a clustering model applicable to time-varying turbulent flows by describing inertial particle clustering in non-equilibrium turbulence as a finite-time relaxation process.

\section{Methods}

The DNS in this study was performed using the Lagrangian Cloud Simulator in Julia (LCS.jl) \cite{tominaga2026LCS}.
LCS.jl is a multi-platform simulation code based on the Lagrangian Cloud Simulator (LCS) by Onishi et al. \cite{onishi2015lagrangian} and shows high scalability for large-scale particle-tracking simulations on GPU environments.

\subsection{Flow Phase}

The carrier fluid follows the incompressible Navier--Stokes equations:
\begin{equation}
\nabla \cdot \vect{U} = 0,
\end{equation}
\begin{equation}
\frac{\partial \vect{U}}{\partial t} + (\vect{U} \cdot \nabla)\vect{U}
= -\frac{1}{\rho}\nabla p + \nu \nabla^2 \vect{U} + \vect{F}(\vect{x},t).
\end{equation}
Here, $\vect{U}$ is the fluid velocity, $\rho$ is the density, $p$ is the pressure, and $\vect{F}$ is the forcing term.
Turbulence was maintained by the forcing term $\vect{F}$, which injects energy into large scales with wavenumber $|\vect{k}| < 2.5$.
Reduced Communication Forcing (RCF), which is a modified Large Scale Forcing method, was used for turbulent forcing \cite{onishi2013efficient}.
RCF enables energy injection only into large scales while reducing communication and computational costs by applying a box mean filter before transformation into wavenumber space, rather than transforming the entire flow field into wavenumber space.
In RCF, the energy injection rate
\begin{equation}
P(t) \equiv \frac{1}{V}\int_V \vect{U}\cdot \vect{F}\,\dd V
\end{equation}
is controlled to a prescribed value by adjusting the amplitude of the forcing term $\vect{F}$.
Here, $V$ is the volume of the computational domain, and $\dd V$ is the volume element.
The governing equations were discretized on a uniform Cartesian grid, and the variables were stored on a MAC arrangement \cite{HarlowWelch1965}.
For spatial discretization, the conservative fourth-order central-difference scheme of Morinishi et al. \cite{morinishi1998fully} was applied to the convective term, a fourth-order central-difference scheme was applied to the viscous term, and a second-order central-difference scheme was applied to the pressure term.
Time integration was performed using a two-stage second-order Runge--Kutta method.
Pressure--velocity coupling was handled using the HSMAC method with Red--Black coloring \cite{hirt1972calculating}.
Iterations were continued until the root-mean-square (RMS) of the velocity divergence became smaller than $\delta/\Delta$.
Here, $\Delta$ is the grid spacing, and $\delta$ was set to $10^{-3}$ following Onishi et al. \cite{onishi2015lagrangian}.
For stability, the time step $\Delta t$ was determined to satisfy the convective and viscous CFL constraints:
\begin{equation}
\mathrm{CFL}_{\mathrm{conv}} = \frac{\max(\|\vect{U}\|)\Delta t}{\Delta} \leq C_1,
\quad
\mathrm{CFL}_{\mathrm{vis}} = \frac{\nu \Delta t}{\Delta^2} \leq C_2.
\end{equation}
Here, $C_1 = C_2 = 0.3$ was used.
The computational domain was a cube with side length $2\pi L_0$, and periodic boundary conditions were applied in all directions.
The grid spacing was $\Delta = 2\pi L_0/N$.
Here, $N$ is the number of grid points in each direction.
The bulk Reynolds number is defined as $\Rey = U_0L_0/\nu$.
Here, $U_0$ is the characteristic velocity, and $L_0$ is the characteristic length.
Three-dimensional domain decomposition was used for parallelization across processes.
The number of divisions was selected so that each subdomain was as close to a cube as possible, and the communication cost was reduced by minimizing the surface-area-to-volume ratio.
Onishi et al. \cite{onishi2013efficient} showed that these numerical methods maintain sufficient accuracy even compared with spectral methods.

\subsection{Particle Phase}

The particle motion is governed by the following system of ordinary differential equations:
\begin{equation}
\frac{\dd \vect{X}}{\dd t} = \vect{V},
\end{equation}
\begin{equation}
\frac{\dd \vect{V}}{\dd t} = -\frac{f}{\tau_{\mathrm{p}}}\{\vect{V} - \vect{U}(\vect{X},t)\}.
\end{equation}
Here, $\vect{X}$ and $\vect{V}$ are the particle position and velocity, respectively, and $\vect{U}(\vect{X},t)$ is the fluid velocity at the particle position.
$\tau_{\mathrm{p}} = (2/9)(\rho_{\mathrm{p}}/\rho)(r_{\mathrm{p}}^2/\nu)$ is the relaxation time of a particle with radius $r_{\mathrm{p}}$ and density $\rho_{\mathrm{p}}$.
The density ratio between the particles and the fluid, $\rho_{\mathrm{p}}/\rho$, was set to $8.43 \times 10^2$, corresponding to the density ratio between water and air under atmospheric conditions at 1013 hPa and 298 K.
The drag correction factor $f$ and particle Reynolds number $\Rey_{\mathrm{p}}$ are defined as follows \cite{rowehenwood1961}:
\begin{equation}
f = 1 + 0.15\Rey_{\mathrm{p}}^{0.687},
\quad
\Rey_{\mathrm{p}} = \frac{2r_{\mathrm{p}}|\vect{V} - \vect{U}(\vect{X})|}{\nu}.
\end{equation}
The fluid velocity $\vect{U}(\vect{X})$ at the particle position $\vect{X}$ was calculated by trilinear interpolation from the cell-center values at the surrounding eight points.
Time integration was performed using a two-stage second-order Runge--Kutta method.
Following Onishi et al. \cite{onishi2015lagrangian}, the number of particles was set to $N_{\mathrm{p}}=N^3/8$.
The Stokes number $\St$, which represents particle inertia, is defined as the ratio of the particle relaxation time to the Kolmogorov time scale $\tau_\eta = (\nu/\eps)^{1/2}$:
\begin{equation}
\St \equiv \frac{\tau_{\mathrm{p}}}{\tau_\eta}.
\end{equation}
Here, $\eps$ is the energy dissipation rate.
The intensity of particle clustering was evaluated using the radial distribution function $g(r=R)$ at the contact distance $R = 2r_{\mathrm{p}}$.
$g(r=R)$ is the particle-pair density near the contact distance normalized by the particle-pair density for a uniform distribution.
In this study, $g(r=R)$ was calculated using particle pairs satisfying $|r - R|/R < 0.01$.
Hereafter, $g \equiv g(r=R)$ is used for simplicity.

\subsection{Unsteady Forcing}

To generate non-equilibrium turbulence, the energy injection rate $P(t)$ prescribed in RCF was changed stepwise in time.
The interval with high $P(t)$ is defined as the strong phase, and the interval with low $P(t)$ is defined as the weak phase.
The energy injection rates in the strong and weak phases are denoted by $P_{\mathrm{s}}$ and $P_{\mathrm{w}}$, respectively.

In periodic unsteady forcing, $P(t)$ is alternately switched between $P_{\mathrm{s}}$ and $P_{\mathrm{w}}$ with period $T$.
In this case, the energy injection rate $P(t)$ is given by
\begin{equation}
P(t) =
\begin{cases}
P_{\mathrm{s}} & (0 \leq \mathrm{mod}(t,T) < T/2),\\
P_{\mathrm{w}} & (T/2 \leq \mathrm{mod}(t,T) < T).
\end{cases}
\end{equation}

In step-once unsteady forcing, the turbulent flow and particle distribution were first developed to a statistically stationary state by steady forcing with $P(t) = P_{\mathrm{s}}$.
The obtained statistically stationary state was used as the initial condition, and $P(t)$ was switched only once from $P_{\mathrm{s}}$ to $P_{\mathrm{w}}$ at $t = 0$.

\subsection{Simulation Cases}\label{sec:simulation-cases}

Table~\ref{tab:simulation-cases} shows the main simulation conditions.
For each unsteady forcing case, steady forcing simulations were performed with time-independent energy injection rates $P = P_{\mathrm{s}}$ and $P = P_{\mathrm{w}}$.
The statistically stationary states obtained from these simulations are defined as the strong reference state and weak reference state, respectively.
For any quantity $X$, the values in the strong reference state and weak reference state are denoted by $X_{\mathrm{s}}$ and $X_{\mathrm{w}}$, respectively.
$\Rey_{\lambda,\mathrm{s}}$ and $\Rey_{\lambda,\mathrm{w}}$ in Table~\ref{tab:simulation-cases} are the Taylor Reynolds numbers in the strong reference state and weak reference state, respectively.
The Taylor Reynolds number is defined as $\Rey_\lambda \equiv u'\lambda/\nu$.
Here, $u'$ is the root-mean-square velocity fluctuation, and $\lambda$ is the Taylor microscale.
The normalized forcing period is defined as $T^* \equiv T/T_{\mathrm{e}}$.
Here, $T_{\mathrm{e}} \equiv L/u'$ is the large-eddy turnover time, and $L$ is the integral length scale.
$T_{\mathrm{e}}$ was evaluated from the turbulent flow obtained by steady forcing with time-independent $P = (P_{\mathrm{s}} + P_{\mathrm{w}})/2$.
$\St_{\mathrm{s}}$ is the Stokes number based on the Kolmogorov time scale in the strong reference state.

The spatial resolution was evaluated using $k_{\max}l_{\eta,\mathrm{s}}$.
Here, $l_{\eta,\mathrm{s}}$ is the Kolmogorov length scale in the strong reference state, and $k_{\max} = N/2$.
Because $l_\eta$ takes the smallest value in the strong reference state, $k_{\max}l_{\eta,\mathrm{s}}$ represents the lower bound of the resolution in each case.
All cases satisfied $k_{\max}l_{\eta,\mathrm{s}} > 1.5$.
This value is comparable to those in existing DNS studies on inertial-particle RDFs and collision statistics \cite{ireland2016part2}.

\begin{table}
\caption{Simulation cases for unsteady forcing.
For each unsteady case, steady reference simulations were performed at $P = P_{\mathrm{s}}$ and $P = P_{\mathrm{w}}$.
$P_{\mathrm{s}}$ and $P_{\mathrm{w}}$ denote the prescribed energy injection rates in the strong and weak phases, respectively.
$T^*$ denotes the forcing period normalized by the large-eddy turnover time.
$\Rey_{\lambda,\mathrm{s}}$ and $\Rey_{\lambda,\mathrm{w}}$ denote the Taylor Reynolds numbers of the strong and weak reference states, respectively.
$k_{\max}l_{\eta,\mathrm{s}}$ and $\St_{\mathrm{s}}$ are evaluated using the strong reference state.}
\label{tab:simulation-cases}
\begin{ruledtabular}
\begin{tabular}{cccccccccc}
Case & Forcing & $N^3$ & $P_{\mathrm{s}}$ & $P_{\mathrm{w}}$ & $T^*$ & $\Rey_{\lambda,\mathrm{s}}$ & $\Rey_{\lambda,\mathrm{w}}$ & $k_{\max}l_{\eta,\mathrm{s}}$ & $\St_{\mathrm{s}}$ \\
\hline
P1 & periodic & $512^3$ & $0.8$ & $0.2$ & $1.5, 3.0, 6.0, 12.0$ & $206$ & $166$ & $1.64$ & $0.25, 1.0, 4.0$ \\
S1 & step-once & $512^3$ & $1.0$ & $0.0625$ & -- & $216$ & $134$ & $1.55$ & $0.25, 0.5, 1.0, 2.0, 4.0$ \\
S2 & step-once & $256^3$ & $1.0$ & $0.0625$ & -- & $134$ & $83.8$ & $1.55$ & $0.25, 0.5, 1.0, 2.0, 4.0$ \\
\end{tabular}
\end{ruledtabular}
\end{table}

\section{Results and Discussion}

\subsection{Non-equilibrium scaling}\label{sec:noneq-scaling}

\begin{figure}
\maybeincludegraphics[width=0.60\textwidth]{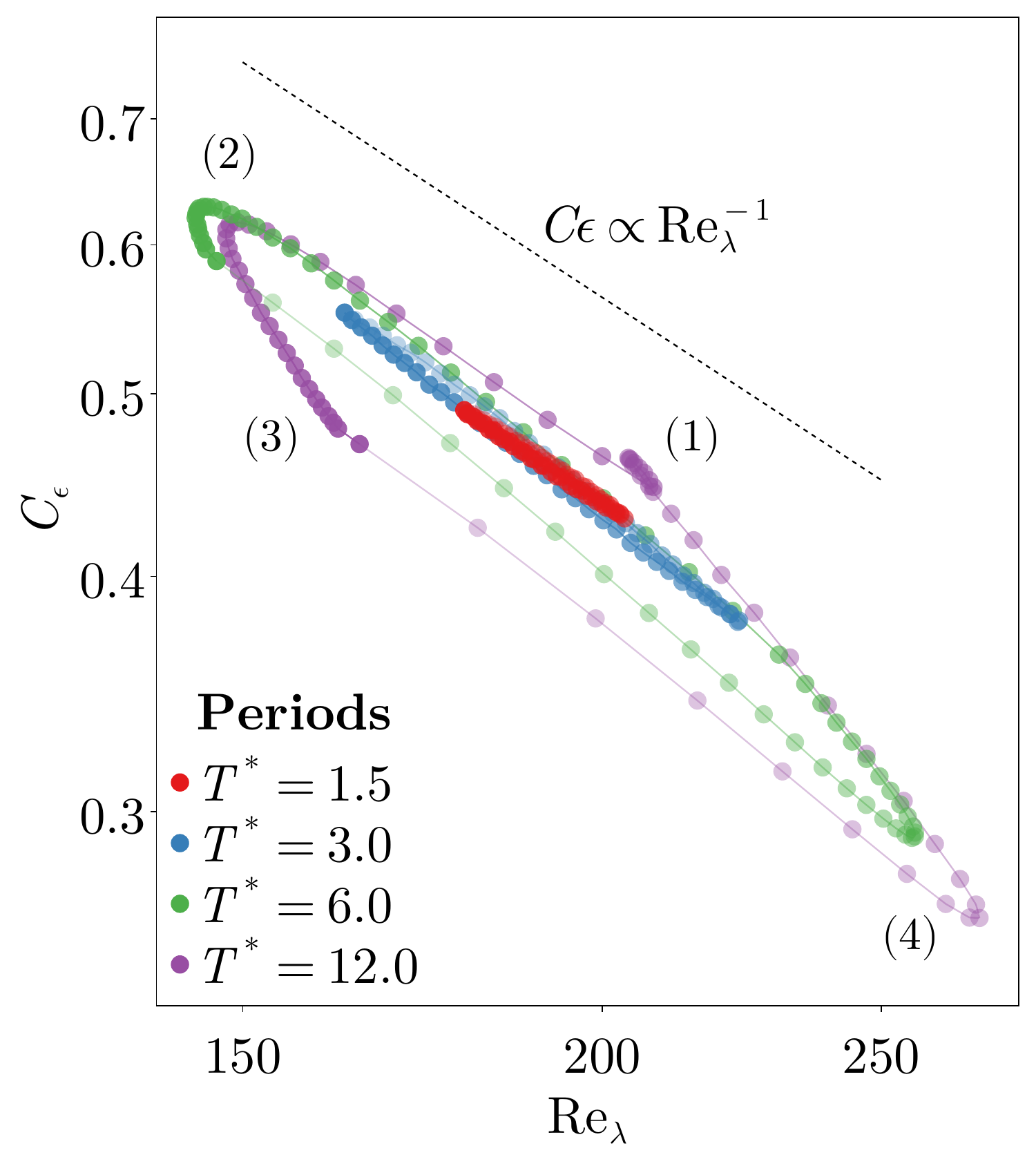}
\caption{$C_\eps$ as a function of $\Rey_\lambda$ for the P1 case in Table~\ref{tab:simulation-cases}.
Color shading from light to dark indicates time within each cycle.
Dashed line indicates the non-equilibrium scaling $C_\eps \propto \Rey_\lambda^{-1}$.}
\label{fig:Re_lambda-vs-C_epsilon}
\end{figure}

Figure~\ref{fig:Re_lambda-vs-C_epsilon} shows $C_\eps$ as a function of $\Rey_\lambda$ for periodic unsteady forcing.
Here, $C_\eps \equiv \eps L/u'^3$.
The simulation condition corresponds to the P1 case in Table~\ref{tab:simulation-cases}.
In P1, phase averaging was performed using time series over 100 forcing cycles.
The maximum relative standard errors of the phase-averaged $C_\eps$ and $\Rey_\lambda$ were less than 2\% for all conditions in the P1 case in Table~\ref{tab:simulation-cases}.

For all forcing periods, the non-equilibrium dissipation scaling \cite{goto2015energy} $C_\eps \propto \Rey_\lambda^{-1}$  was observed.
For $T^* = 12.0$, four states are identified.
States (1) and (3) correspond to the strong and weak phases, respectively, and show $C_\eps \approx 0.5$, which is characteristic of equilibrium turbulence.
States (2) and (4) are transition phases and show non-equilibrium dissipation scaling.
As the forcing period decreased, the turbulence became unable to follow the variation in the energy injection rate.
As a result, the energy injection rate changed before the turbulence was fully developed or decayed, the variation in $\Rey_\lambda$ became smaller, and the hysteresis loop shrank.

This result demonstrates that the unsteady forcing used in this study can systematically generate a non-equilibrium turbulent flow.

\subsection{Hysteresis of Clustering Intensity}\label{sec:hysteresis}

\begin{figure}
\maybeincludegraphics[width=0.60\textwidth]{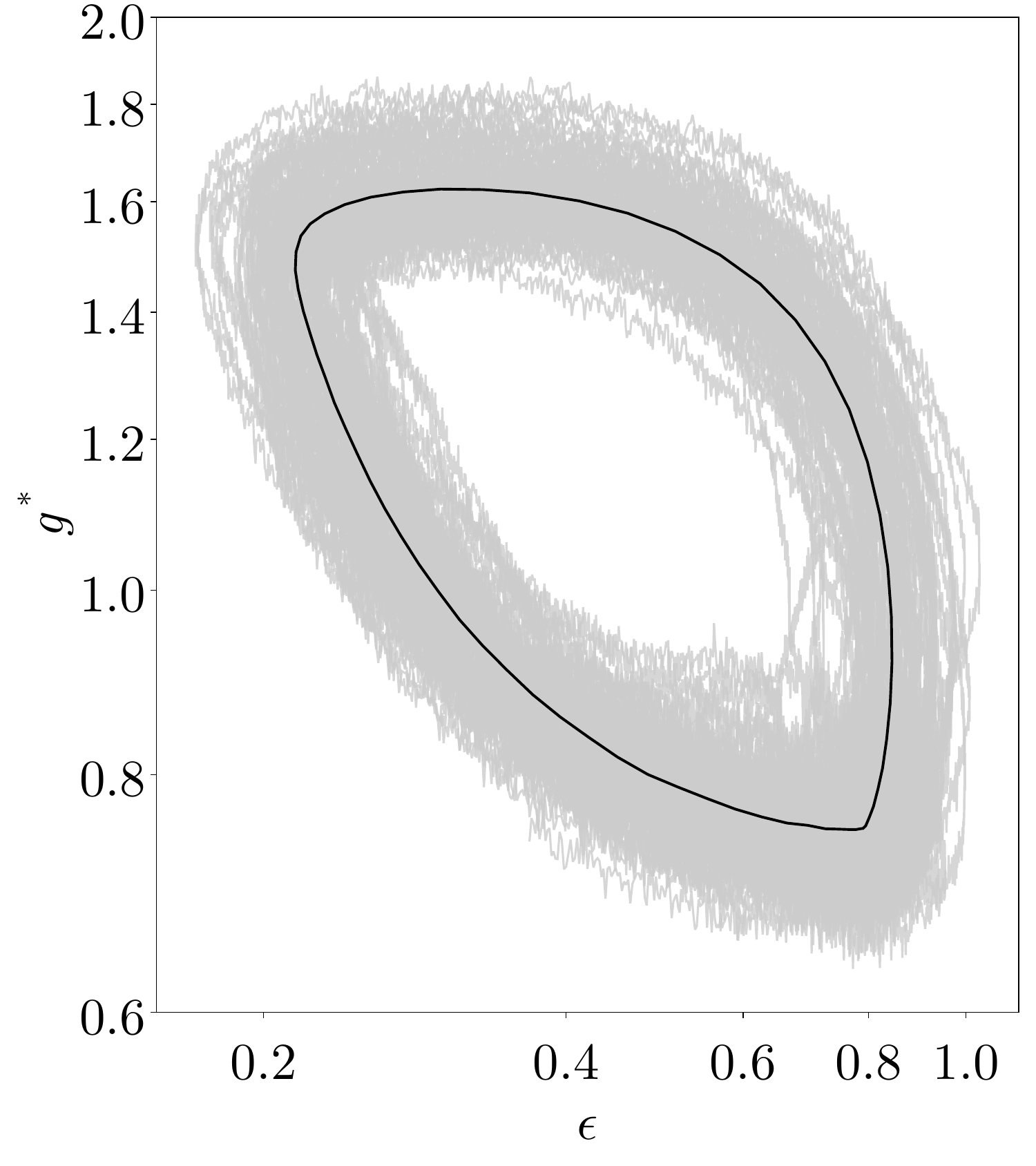}
\caption{Raw and phase-averaged trajectories of $g^*$ as a function of $\eps$ for the P1 case with $\St_\mathrm{s}=4.0$ and $T^*=12.0$ in Table~\ref{tab:simulation-cases}.
Gray lines indicate raw trajectories for individual forcing cycles.
Black line indicates the phase-averaged trajectory.}
\label{fig:hysteresis-fluctuation}
\end{figure}

Figure~\ref{fig:hysteresis-fluctuation} shows $g^*$ as a function of $\eps$ for periodic unsteady forcing.
Here, $g^* \equiv g(t)/\langle g_{\mathrm{mid}}(t)\rangle$.
$g_{\mathrm{mid}}(t)$ is the time series of $g$ in steady forcing with $P = (P_{\mathrm{s}} + P_{\mathrm{w}})/2$, and $\langle g_{\mathrm{mid}}(t)\rangle$ is its time average.
Gray lines indicate raw trajectories for individual cycles.
The black line indicates the phase-averaged trajectory.
The simulation condition corresponds to the P1 case with $\St_\mathrm{s}=4.0$ and $T^*=12.0$ in Table~\ref{tab:simulation-cases}.

The raw trajectories contain large turbulent fluctuations.
Therefore, it is difficult to determine systematic hysteresis in the $\eps$--$g^*$ relationship from the raw trajectories alone.
In P1, phase averaging was therefore performed using time series over 100 forcing cycles.
Phase averaging reduced turbulent fluctuations that are not synchronized with the forcing cycle and extracted the hysteresis loop synchronized with the forcing cycle.
The maximum relative standard error of the phase-averaged $g^*$ was less than 2\% for all conditions in the P1 case.
Below, the dependence of hysteresis on the forcing period and $\St_{\mathrm{s}}$ is discussed based on the phase-averaged trajectories.

\begin{figure}
\centering
\begin{minipage}{0.48\textwidth}
\centering
\maybeincludegraphics[width=\linewidth]{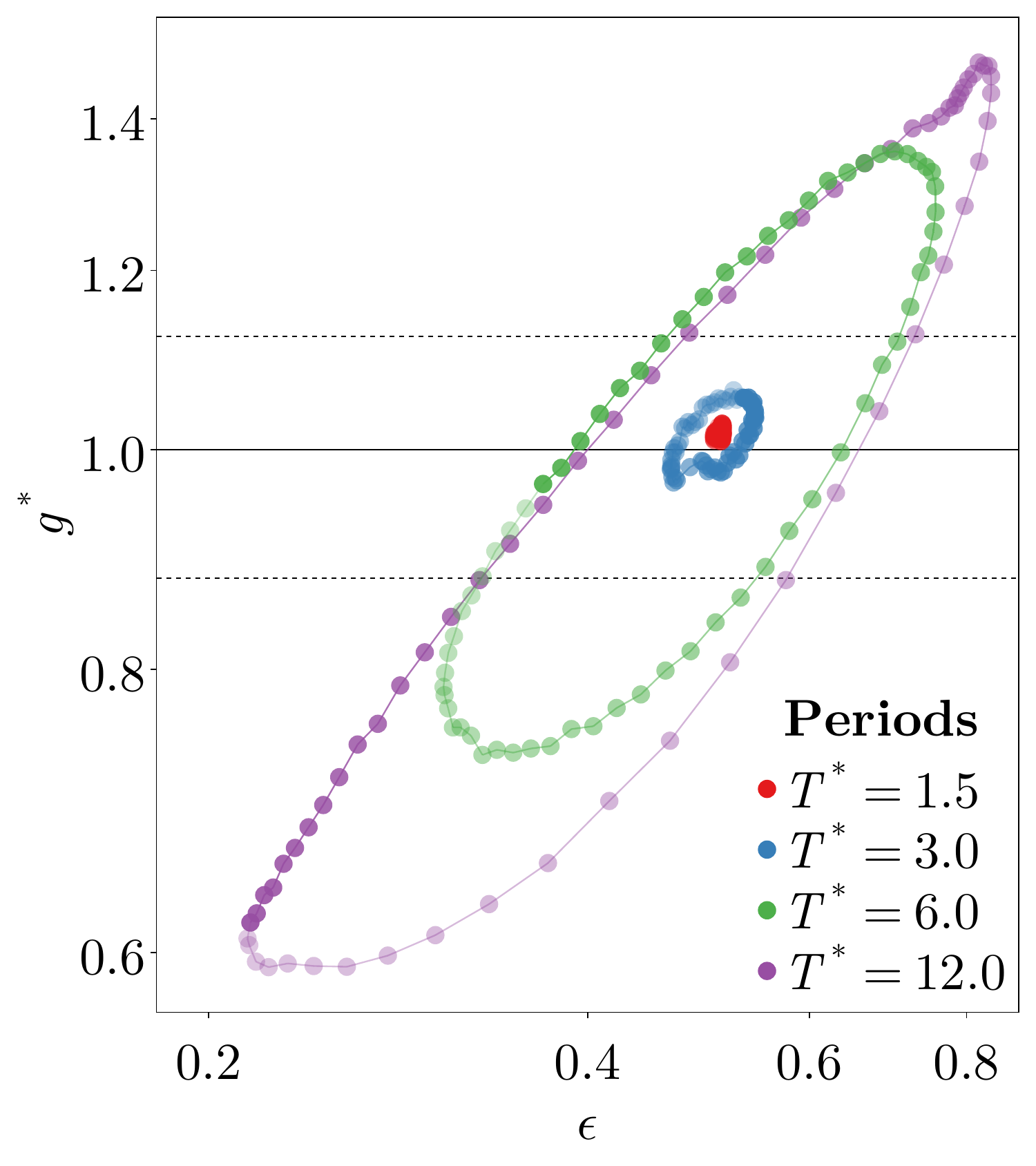}
\vspace{-0.5em}
\centerline{(a) $\St_{\mathrm{s}} = 0.25$}
\end{minipage}\hfill
\begin{minipage}{0.48\textwidth}
\centering
\maybeincludegraphics[width=\linewidth]{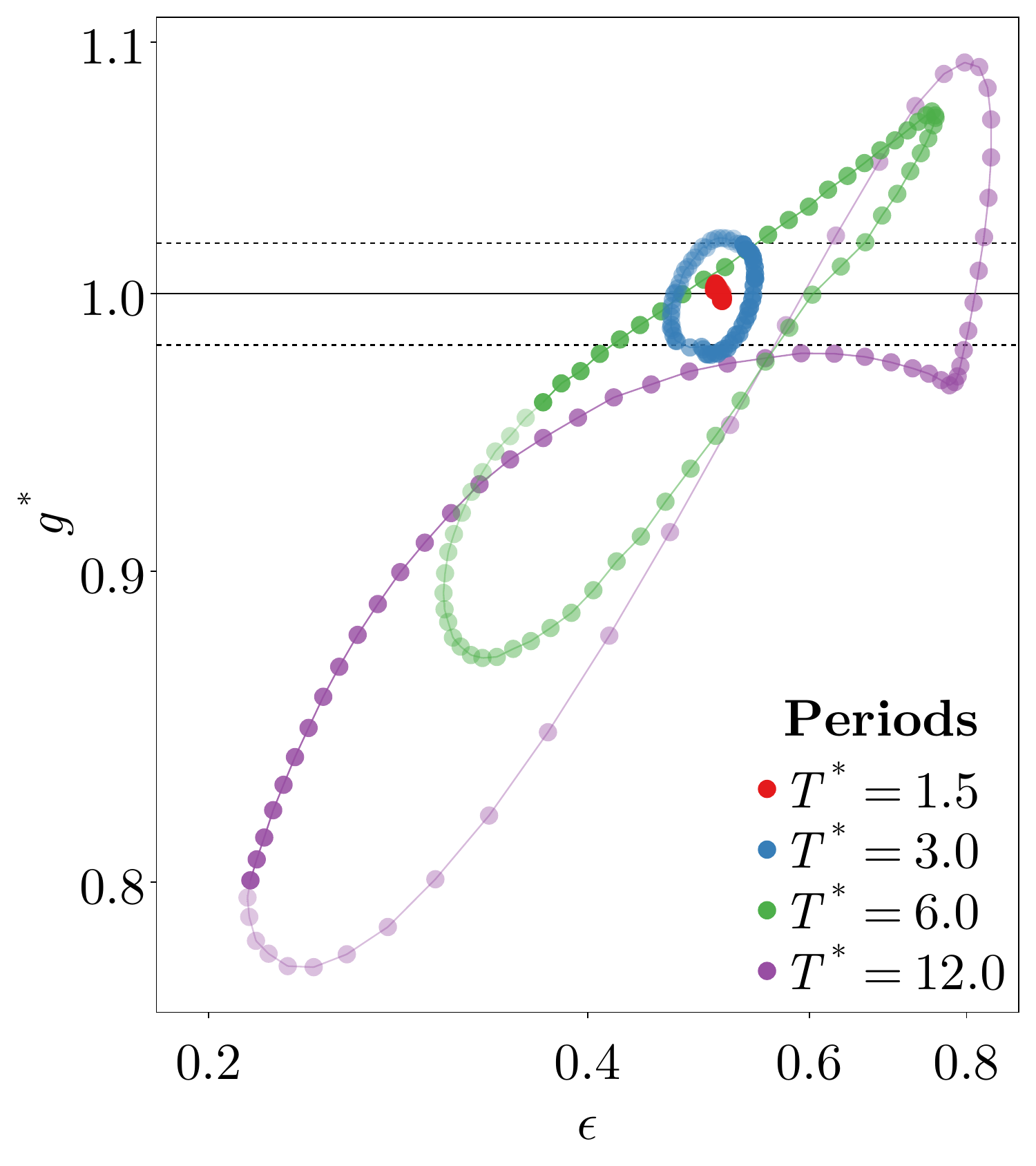}
\vspace{-0.5em}
\centerline{(b) $\St_{\mathrm{s}} = 1.0$}
\end{minipage}

\vspace{0.8em}

\begin{minipage}{0.48\textwidth}
\centering
\maybeincludegraphics[width=\linewidth]{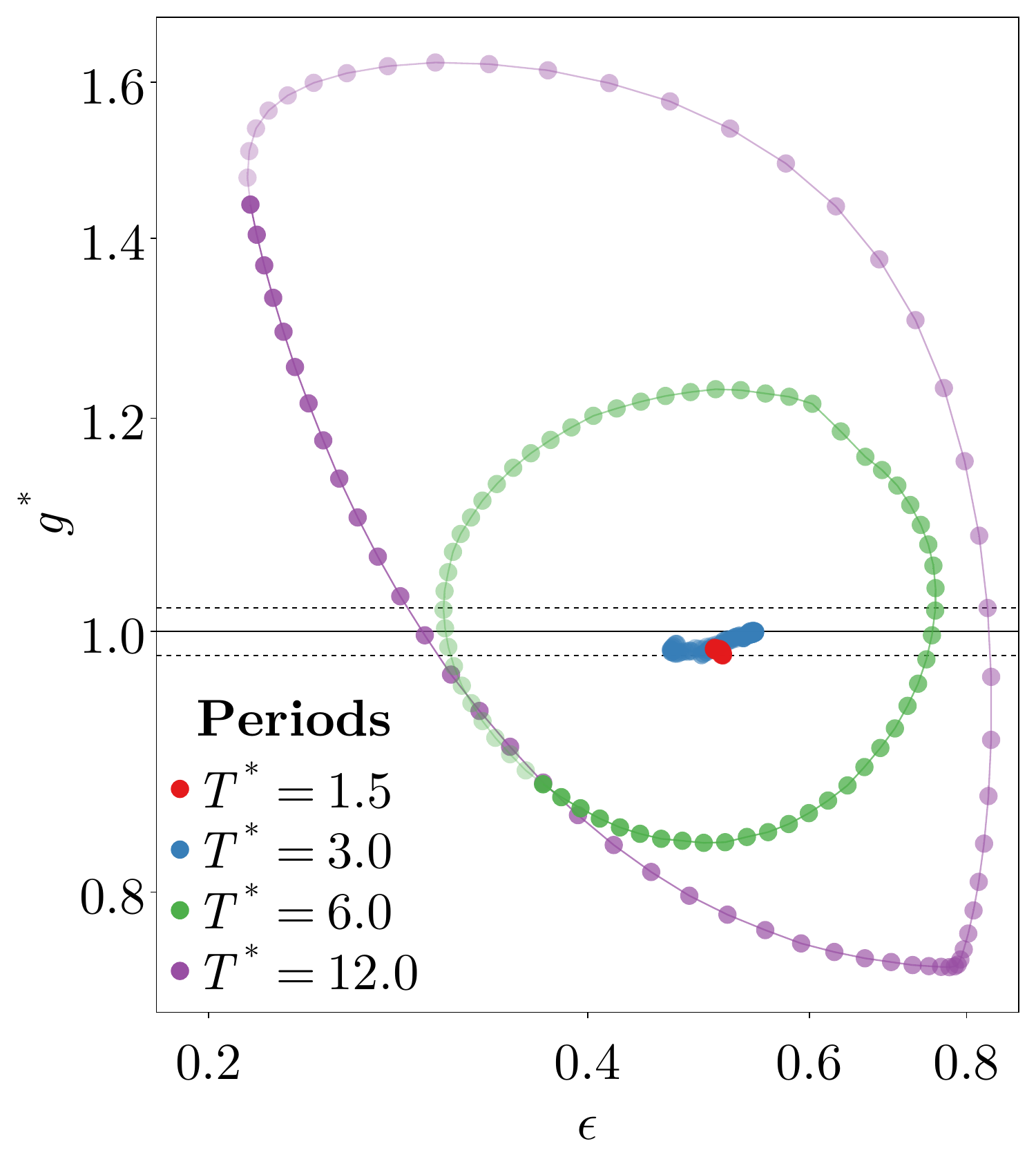}
\vspace{-0.5em}
\centerline{(c) $\St_{\mathrm{s}} = 4.0$}
\end{minipage}
\caption{Phase-averaged $g^*$ as a function of phase-averaged $\eps$ for the P1 case in Table~\ref{tab:simulation-cases}.
Panels (a)--(c) show the results for $\St_{\mathrm{s}} = 0.25$, 1.0, and 4.0, respectively.
Markers represent four normalized forcing periods $T^*$.
Color shading from light to dark indicates time within each cycle.
The solid horizontal line indicates $g^* = 1$.
The dashed horizontal lines indicate $1 \pm \sigma$.}
\label{fig:eps-vs-gr}
\end{figure}

Figure~\ref{fig:eps-vs-gr} shows the phase-averaged $g^*$ as a function of the phase-averaged $\eps$ for periodic unsteady forcing.
The simulation condition corresponds to the P1 case in Table~\ref{tab:simulation-cases}.
The horizontal dashed lines indicate $1 \pm \sigma$.
Here, $\sigma$ is the standard deviation of $g_{\mathrm{mid}}(t)/\langle g_{\mathrm{mid}}(t)\rangle$ and provides a reference fluctuation scale of $g^*$ in statistically stationary turbulence.

For all $\St_{\mathrm{s}}$, the largest hysteresis loop was formed at $T^* = 12.0$.
In particular, for $\St_{\mathrm{s}} = 4.0$ and $T^* = 12.0$, $g^*(t)$ took two different values, 0.80 and 1.56, depending on the flow history for the same $\eps(t) = 0.5$.
As the forcing period decreased, the hysteresis loop shrank.
For $T^* = 6.0$ and $T^* = 12.0$, clear hysteresis loops extending beyond the reference fluctuation band $[1 - \sigma, 1 + \sigma]$ were observed in the $\eps$--$g^*$ plane.
For $T^* = 3.0$, the loop was mostly within this band, and the hysteresis remained comparable to the fluctuation scale of $g^*$.
For $T^* = 1.5$, $\eps$ and $g^*$ were almost constant, and no hysteresis was observed.

The $\St_{\mathrm{s}}$ dependence of the loop shape can be interpreted from the relationship between the normalized equilibrium values $g_{\mathrm{s}}^*$ and $g_{\mathrm{w}}^*$, which correspond to the strong reference state and weak reference state, respectively.
Here, $g_{\mathrm{s}}$ and $g_{\mathrm{w}}$ are the RDFs in the strong reference state and weak reference state, respectively.
In addition, $g_{\mathrm{s}}^* \equiv g_{\mathrm{s}}/\langle g_{\mathrm{mid}}\rangle$ and $g_{\mathrm{w}}^* \equiv g_{\mathrm{w}}/\langle g_{\mathrm{mid}}\rangle$.
In the present simulations, the particle radius was fixed in each case.
Therefore, changes in $\eps$ change the Kolmogorov time scale and the instantaneous Stokes number $\St(t)$.
As a result, the relationship between $g_{\mathrm{s}}^*$ and $g_{\mathrm{w}}^*$ characterizes the direction of the trajectory in the $\eps$--$g^*$ plane.

For $\St_{\mathrm{s}} = 0.25$, $g_{\mathrm{s}}^* > g_{\mathrm{w}}^*$.
Therefore, $\eps$ and $g^*$ are generally positively correlated, and the loop in Fig.~\ref{fig:eps-vs-gr}(a) has an upward-sloping shape.
In contrast, for $\St_{\mathrm{s}} = 4.0$, $g_{\mathrm{s}}^* < g_{\mathrm{w}}^*$.
Therefore, $\eps$ and $g^*$ are generally negatively correlated, and the loop in Fig.~\ref{fig:eps-vs-gr}(c) has a downward-sloping shape.
For $\St_{\mathrm{s}} = 1.0$, changes in $\eps$ cause $\St(t)$ to pass near the maximum of the equilibrium clustering intensity.
In this case, the direction of change in the equilibrium value with increasing $\eps$ can reverse within the same cycle.
Therefore, the $\eps$--$g^*$ trajectory in Fig.~\ref{fig:eps-vs-gr}(b) has a non-monotonic shape with self-crossing.
Thus, for $\St_{\mathrm{s}} = 1.0$, the relationship between $\eps$ and $g^*$ cannot be described as a simple positive or negative correlation.

This result demonstrates that, when the forcing period exceeds several large-eddy turnover times, the instantaneous-equilibrium approximation is not appropriate for describing clustering intensity in non-equilibrium turbulence.

\subsection{Linear Relaxation Model}\label{sec:relaxation-model}

\begin{figure}
\maybeincludegraphics[width=0.60\textwidth]{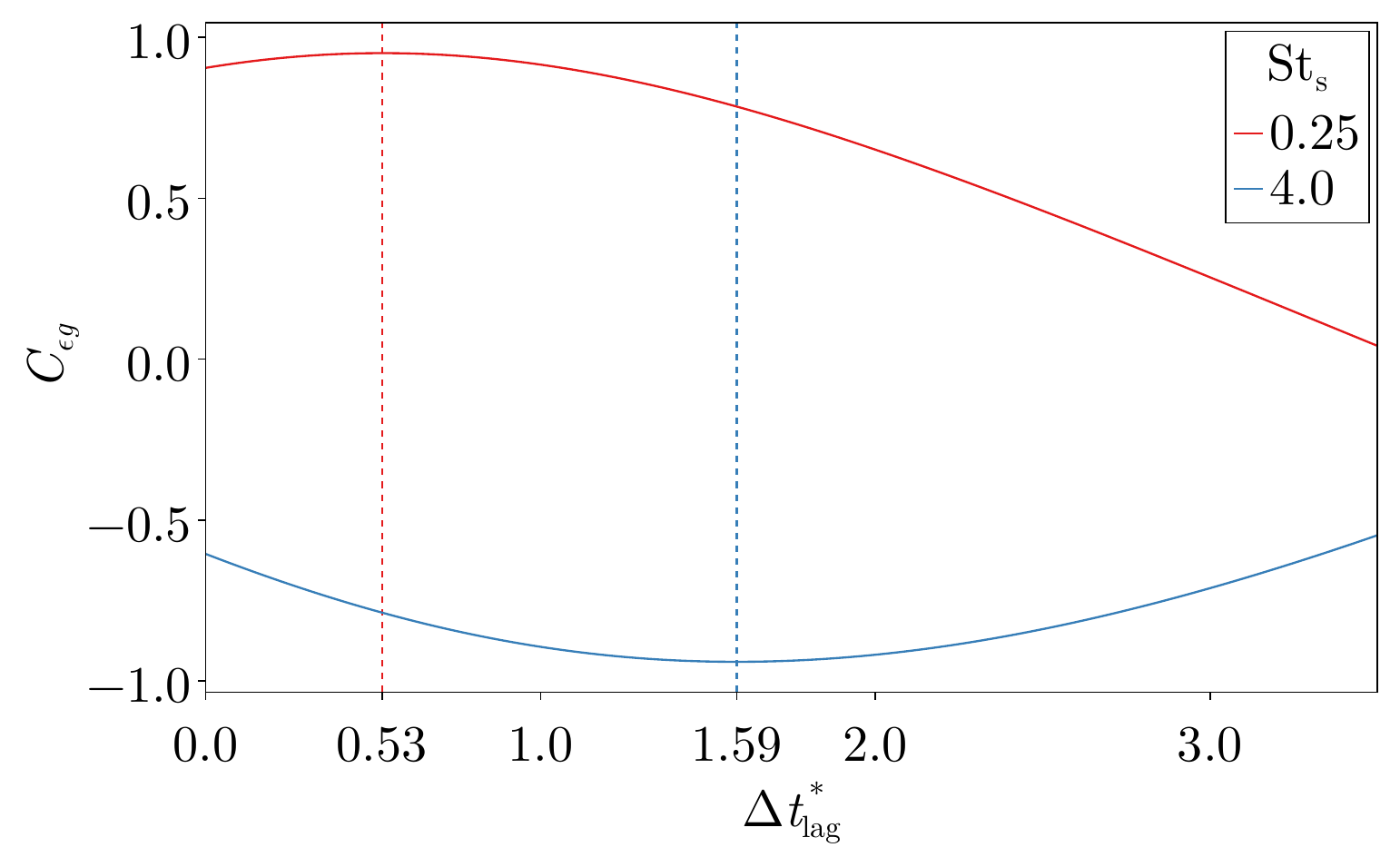}
\caption{Cross-correlation function $C_{\eps g}(\Delta t_{\mathrm{lag}}^*)$ between $\eps(t)$ and $g(t)$ as a function of the normalized lag time $\Delta t_{\mathrm{lag}}^*$ for the P1 case with $T^* = 12.0$ in Table~\ref{tab:simulation-cases}.
Red and blue lines indicate $\St_{\mathrm{s}} = 0.25$ and $\St_{\mathrm{s}} = 4.0$, respectively.
Vertical dashed lines indicate the lag time $\tau_g^*$ that maximizes $|C_{\eps g}(\Delta t_{\mathrm{lag}}^*)|$.}
\label{fig:relaxation-time-estimation}
\end{figure}

Figure~\ref{fig:relaxation-time-estimation} shows the cross-correlation function $C_{\eps g}$ as a function of the normalized lag time $\Delta t_{\mathrm{lag}}^*$ for periodic unsteady forcing.
The simulation condition corresponds to the P1 case in Table~\ref{tab:simulation-cases}.
Here, the time series for $T^* = 12.0$ was used because hysteresis was most pronounced and the phase difference between $\eps(t)$ and $g(t)$ was easy to identify.

To quantify the response time of clustering intensity to the forcing variation, the normalized cross-correlation function between $\eps(t)$ and $g(t)$ was calculated as
\begin{equation}
C_{\eps g}(\Delta t_{\mathrm{lag}})
=
\frac{\langle \eps'(t)g'(t+\Delta t_{\mathrm{lag}})\rangle}{\sigma_\eps \sigma_g}.
\end{equation}
Here, $\eps'$ and $g'$ are the signals with their means removed, and $\sigma_\eps$ and $\sigma_g$ are their standard deviations.
The normalized lag time is defined as $\Delta t_{\mathrm{lag}}^* \equiv \Delta t_{\mathrm{lag}}/T_{\mathrm{e}}$ using the large-eddy turnover time $T_{\mathrm{e}}$ evaluated from steady forcing with time-independent $P = (P_{\mathrm{s}} + P_{\mathrm{w}})/2$.
The clustering relaxation time is denoted by $\tau_g$, and its normalized value is defined as $\tau_g^* \equiv \tau_g/T_{\mathrm{e}}$.
Because the response sign depends on $\St_{\mathrm{s}}$, $\tau_g^*$ was defined as the positive lag time $\Delta t_{\mathrm{lag}}^*$ that maximizes $|C_{\eps g}(\Delta t_{\mathrm{lag}}^*)|$.

For $\St_{\mathrm{s}} = 1.0$, the $\eps$--$g^*$ relationship has a non-monotonic shape with self-crossing, as shown in Fig.~\ref{fig:eps-vs-gr}(b).
Therefore, it is difficult to interpret the peak of the cross-correlation function as a single linear relaxation time, and this case was excluded from the cross-correlation analysis.
$\tau_g^*$ increased with the Stokes number and was estimated as $\tau_g^* \approx 0.53$ and $\tau_g^* \approx 1.59$ for $\St_{\mathrm{s}} = 0.25$ and $4.0$, respectively.

This result suggests that $\tau_g$ has a time scale comparable to $T_{\mathrm{e}}$ and increases with $\St$.
Therefore, the following linear relaxation model is proposed:
\begin{equation}
\frac{\dd g_{\mathrm{noneq}}}{\dd t}
= \frac{g_{\mathrm{eq}}(t)-g_{\mathrm{noneq}}(t)}{\tau_g},
\quad
\tau_g = C T_{\mathrm{e}}(t)\St(t)^a.
\label{eq:model}
\end{equation}

Here, instantaneous values were used for $T_{\mathrm{e}}(t)$ and $\St(t)$, where $T_{\mathrm{e}}(t)$ was evaluated as $L(t)/u'(t)$.
$g_{\mathrm{eq}}(t)$ was obtained from the model of Onishi et al. \cite{onishi2016reynolds} using the instantaneous $\Rey_\lambda(t)$ and $\St(t)$.
However, the prediction of Onishi et al. \cite{onishi2016reynolds} did not completely agree with the statistically stationary values in the present DNS.
Therefore, $g_{\mathrm{eq}}(t)$ was linearly scaled so that it agreed with the DNS values of $g$ in the statistically stationary states before and after switching at $t = 0$ and $t \to \infty$, respectively.
This scaling used only the statistically stationary values before and after switching and did not use the time-series information of $g(t)$ during the transient response.
This procedure made the eq-model and noneq-model share the same statistically stationary endpoints.
Thus, the difference between the two models mainly corresponds to the presence or absence of finite-time relaxation.
The model assuming instantaneous equilibrium is referred to as the eq-model, and the linear relaxation model given by Eq.~\eqref{eq:model} is referred to as the noneq-model.

To identify the parameters in the linear relaxation model, the S1 case in Table~\ref{tab:simulation-cases} was used.
$C$ and $a$ were determined by minimizing the mean time-series error between $g_{\mathrm{noneq}}(t)$ obtained by numerically solving Eq.~\eqref{eq:model} and the DNS value of $g(t)$.
As a result, $C = 1.0$ and $a = 0.40$ were obtained.
For example, at $t = 0$, where $\St(0) = \St_{\mathrm{s}}$, $\tau_g(0)/T_{\mathrm{e}}(0) = \St(0)^{0.40} = \St_{\mathrm{s}}^{0.40} \approx 0.58$ and $1.72$ for $\St_{\mathrm{s}} = 0.25$ and $4.0$, respectively.
These values are close to $\tau_g^* \approx 0.53$ and $1.59$ estimated from the cross-correlation analysis.
This consistency supports the interpretation that the observed hysteresis originates from a finite relaxation time of clustering intensity.

\begin{figure}
\maybeincludegraphics[width=0.80\textwidth]{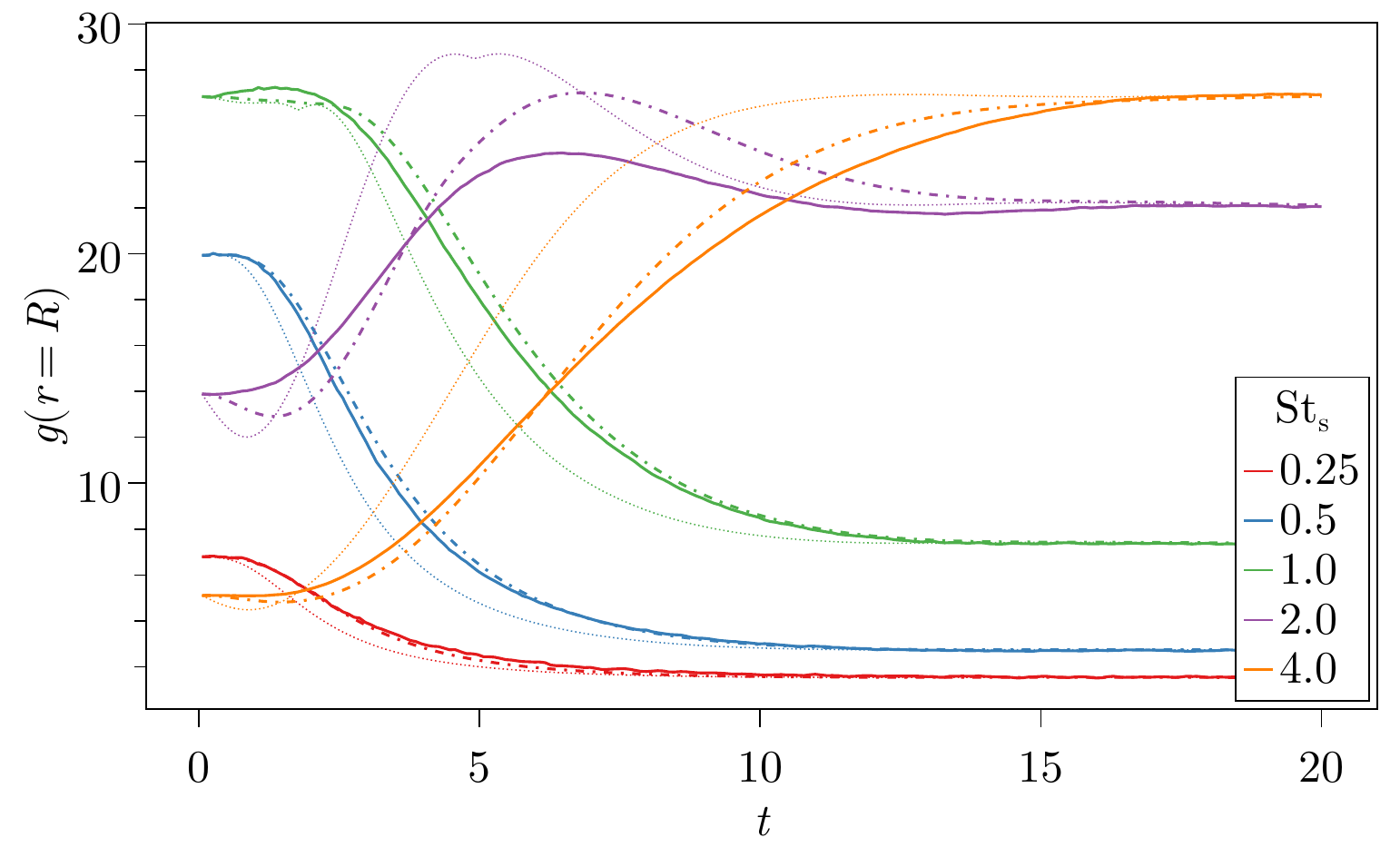}
\caption{Time evolution of $g(r=R)$ for the S1 case in Table~\ref{tab:simulation-cases}.
Solid, dotted, and dash-dotted lines indicate DNS, the eq-model, and the noneq-model, respectively.
Colors indicate $\St_{\mathrm{s}}$ in $\{0.25, 0.5, 1.0, 2.0, 4.0\}$.}
\label{fig:model-noneq}
\end{figure}

Figure~\ref{fig:model-noneq} shows the time evolution of $g$ for DNS, the eq-model, and the noneq-model.
The simulation condition corresponds to the S1 case in Table~\ref{tab:simulation-cases}.
In S1, ensemble averaging was performed using 256 cases.
The maximum relative standard error of $g$ was less than 2\% for all Stokes numbers.
For all Stokes numbers, the noneq-model reproduced the DNS data better than the eq-model.
In contrast, for $\St_{\mathrm{s}} = 2.0$, both the eq-model and noneq-model overestimated the maximum value of $g$.
This overestimation can be interpreted as follows.
Because the eq-model has no relaxation time, $g_{\mathrm{eq}}$ instantaneously follows the time variation in $\St(t)$ and reaches a maximum near $\St \sim 1$.
In contrast, in the DNS, $g$ does not reach the maximum value corresponding to $\St \sim 1$ because $\St(t)$ remains near $\St \approx 1$ only for a short time.
Because the noneq-model uses $g_{\mathrm{eq}}$ from the eq-model as input, it is affected by this overestimated peak.

Table~\ref{tab:model-error} shows the maximum relative error $E_\mathrm{model}$ of the eq-model and noneq-model for the S1 and S2 cases.
The maximum relative error $E_\mathrm{model}$ is defined as
\begin{equation}
E_{\mathrm{model}} = \max_t\left|\frac{g_{\mathrm{model}}(t)-g(t)}{g(t)}\right|.
\end{equation}
The S2 case is an independent case with a different Taylor Reynolds number, while using the same step-once forcing and the same $P_{\mathrm{s}}/P_{\mathrm{w}}$ ratio as the S1 case.
The S2 case was used to validate whether the relaxation time scaling identified from the S1 case is also effective for a different Reynolds-number condition.
In S2, ensemble averaging was also performed using 256 cases.
The maximum relative standard error of $g$ was less than 2\% for all Stokes numbers.

For all Stokes numbers in the S1 and S2 cases, the noneq-model showed smaller errors than the eq-model.
In particular, for $\St_{\mathrm{s}} = 4.0$, the error was 10\% for the noneq-model, whereas it was 49\% for the eq-model in the S1 case.
In the S2 case, the error was 22\% for the noneq-model, whereas it was 76\% for the eq-model.
The relative error reduction was smaller for small Stokes numbers ($\St_{\mathrm{s}} = 0.25$ and $0.5$) because $\tau_g^*$ was relatively short and the instantaneous-equilibrium approximation was more likely to hold approximately.

These results demonstrate that accounting for finite-time relaxation is effective for improving the prediction accuracy of clustering intensity in non-equilibrium turbulence.

\begin{table}
\caption{Maximum relative error $E_\mathrm{model}$ of the eq-model and noneq-model for the S1 and S2 cases in Table~\ref{tab:simulation-cases}.}
\label{tab:model-error}
\begin{ruledtabular}
\begin{tabular}{ccccc}
$\St_{\mathrm{s}}$ & $E_{\mathrm{eq}}$ (S1) & $E_{\mathrm{noneq}}$ (S1) & $E_{\mathrm{eq}}$ (S2) & $E_{\mathrm{noneq}}$ (S2) \\
\hline
0.25 & 23\% & 12\% & 34\% & 27\% \\
0.5 & 24\% & 7.7\% & 38\% & 18\% \\
1.0 & 21\% & 6.2\% & 28\% & 6.6\% \\
2.0 & 32\% & 12\% & 34\% & 14\% \\
4.0 & 49\% & 10\% & 76\% & 22\% \\
\end{tabular}
\end{ruledtabular}
\end{table}

\section{Conclusions}

In this study, the temporal response of inertial particle clustering in non-equilibrium turbulence is investigated using DNS of homogeneous isotropic turbulence with unsteady forcing, in which the energy injection rate varies in time.

First, the periodic responses of the turbulent flow and clustering intensity were systematically evaluated by varying the forcing period.
The turbulent flow showed non-equilibrium dissipation scaling, $C_\eps \propto \Rey_\lambda^{-1}$, for all $T^*$.
This result demonstrates that the unsteady forcing used in this study can systematically generate a non-equilibrium turbulent flow.
The relationship between $\eps(t)$ and $g^*(t)$ showed hysteresis.
However, systematic evaluation of the hysteresis is difficult because the raw trajectories contain turbulent fluctuations that are not synchronized with the forcing variation.
Therefore, phase averaging was used to reduce turbulent fluctuations that are not synchronized with the forcing variation and to extract the response synchronized with the forcing variation.
The relationship between the obtained phase-averaged $g^*(t)$ and phase-averaged $\eps(t)$ showed hysteresis exceeding the band $[1 - \sigma, 1 + \sigma]$, which represents the reference fluctuation scale in statistically stationary turbulence, for $T^* = 6.0$ and $T^* = 12.0$ for all $\St_{\mathrm{s}}$ considered here.
In particular, for $\St_{\mathrm{s}} = 4.0$ and $T^* = 12.0$, $g^*(t)$ took two different values, 0.80 and 1.56, depending on the flow history for the same $\eps(t) = 0.5$.
This result demonstrates that, when the forcing period exceeds several large-eddy turnover times, the instantaneous-equilibrium approximation is not appropriate for describing clustering intensity in non-equilibrium turbulence.

Next, a linear relaxation model was constructed using transient responses, in which clustering intensity approaches the instantaneous-equilibrium value with a finite relaxation time.
For the relaxation time, the scaling $\tau_g = 1.0 T_{\mathrm{e}}(t)\St(t)^{0.40}$, which depends on the instantaneous large-eddy turnover time $T_{\mathrm{e}}(t)$ and Stokes number $\St(t)$, was identified.
The linear relaxation model showed smaller errors than the model assuming instantaneous equilibrium for all $\St_{\mathrm{s}}$ conditions.
In particular, for $\St_{\mathrm{s}} = 4.0$, the maximum relative error was reduced from 49\% to 10\%.
In an independent validation case with a different Reynolds number, the maximum relative error was reduced from 76\% to 22\%.
These results demonstrate that accounting for finite-time relaxation is effective for improving the prediction accuracy of clustering intensity in non-equilibrium turbulence.

This study provides a basis for a clustering model applicable to time-varying turbulent flows by describing inertial particle clustering in non-equilibrium turbulence as a finite-time relaxation process.

\begin{acknowledgments}
This work used computational resources of the TSUBAME4.0 supercomputer provided by Institute of Science Tokyo through the Joint Usage/Research Center for Interdisciplinary Large-scale Information Infrastructures (JHPCN) and the High Performance Computing Infrastructure (HPCI) in Japan (Project ID: jh240041).
\end{acknowledgments}

\section*{AUTHOR DECLARATIONS}

\subsection*{Conflict of Interest}

The authors have no conflicts to disclose.

\subsection*{Author Contributions}

Taketo Tominaga: Conceptualization (lead); Data curation (lead); Formal analysis (lead); Investigation (lead); Methodology (lead); Software (lead); Validation (lead); Visualization (lead); Writing -- original draft (lead); Writing -- review and editing (equal).

Ryo Onishi: Conceptualization (supporting); Funding acquisition (lead); Methodology (supporting); Resources (lead); Software (supporting); Supervision (lead); Validation (supporting); Writing -- review and editing (equal).

\section*{DATA AVAILABILITY STATEMENT}

The data that support the findings of this study are available from the corresponding author upon reasonable request.

\bibliography{refs}

\end{document}